\documentclass[showpacs,preprintnumbers,amsmath,amssymb,pre,superscriptaddress]{revtex4-1}

\usepackage{graphicx}
\usepackage{color}      
\usepackage{longtable}

\begin{document}

\title[Statistics of the stochastically-forced Lorenz attractor
by the Fokker-Planck and cumulant equations]{Statistics of the stochastically-forced Lorenz attractor
by the Fokker-Planck equation and cumulant expansions}

\author{Altan Allawala} 
\email{allawala@brown.edu}

\author{J. B. Marston}
\email{marston@brown.edu}

\address{Department of Physics, Box 1843, Brown University, Providence, Rhode Island 02912-1893, USA}

\begin{abstract}
We investigate the Fokker-Planck description of the equal-time statistics of the three-dimensional Lorenz-63 attractor with additive white noise.  The invariant measure is found by computing the zero (or null) mode of the linear Fokker-Planck operator as a problem of sparse linear algebra.  Two variants are studied: A self-adjoint construction of the linear operator, and the replacement of diffusion with hyperdiffusion. 
We also access the low-order statistics of the system by a perturbative expansion in equal-time cumulants.  Comparison is made to statistics obtained by the standard approach of accumulation via direct numerical simulation. Theoretical and computational aspects of the Fokker-Planck and cumulant expansion methods are discussed.
\end{abstract}


\maketitle

\section{Introduction}
\label{Introduction}

Chaotic dynamical systems often have a well-defined statistical steady state. 
Traditionally statistics are estimated by 
their accumulation through Direct Numerical Simulation (DNS) starting from 
an ensemble of initial conditions.  If the basin of attraction 
is ergodic, ensemble averaging can be replaced by time-averaging over a single
long trajectory.  Rare but large deviations may occur, however, necessitating 
extremely long integration times.  An alternative and more efficient approach
solves for the statistics directly.  Depending on the question to be answered, such Direct Statistical Simulation (DSS) 
can focus on various statistical quantities such as the the probability distribution function (PDF) or invariant measure, 
the low-order equal-time moments, autocorrelations in time, or large deviations.

This paper presents two different types of DSS. The first, the Fokker-Planck equation (FPE), describes the flow of probability density in phase space, respecting the conservation of total probability.
Consider a trajectory governed by the differential equation:
\begin{equation}
\frac{d\vec{x}}{dt} = \vec{V}(\vec{x}) + \vec{\eta}(t)
\end{equation}
where $\vec{\eta}(t)$ is additive stochastic forcing. The FPE for this
system is:
\begin{equation}
\frac{\partial P(\vec{x}, t)}{\partial t} = -\hat{L}_{FPE} P(\vec{x}, t)
\label{FPE}
\end{equation} 
where we will call $\hat{L}_{FPE}$ the (linear) FPE operator. The placement of the negative sign in front of the operator $\hat{L}_{FPE}$ in Eq. \ref{FPE} is for convenience:  The operator $\hat{L}_{FPE}$ is then semipositive definite, and the steady-state statistics are determined by the ground state.  
In the special case where $\vec{\eta}$ is Gaussian
additive white noise with no mean and covariance given by 
\begin{equation}
\langle \eta_{i}(t)~ \eta_{j}(t^\prime) \rangle = 2 \Gamma_{ij}~ \delta(t-t^\prime),
\label{stochastic}
\end{equation}
with angular brackets $\langle \ldots \rangle$ indicating a short time-average,
only a finite number of terms appear in the FPE operator \cite{pawula1967approximation}:
\begin{equation}
\hat{L}_{FPE} P =  \vec{\nabla} \cdot (\vec{V} P) - \Gamma \nabla^{2} P.
\label{LFPE}
\end{equation}
Additive stochastic forcing smears out the PDF $P$
through diffusion in phase space.  A canonical example
is the one-dimensional Ornstein-Uhlenbeck process with trajectories
governed by 
\begin{equation}
\dot{x} = -a x + \eta(t).
\end{equation}
The corresponding FPE is
\begin{equation}
\frac{\partial P(x, t)}{\partial t} = \frac{\partial}{\partial x} [ a x P(x, t)] + \Gamma \frac{\partial^{2}P(x, t)}{\partial x^{2}}
\end{equation}
that has a steady-state solution that is readily found to be Gaussian:
\begin{equation}
P(x) = \sqrt{\frac{a}{2\pi\Gamma}}~ e^{-ax^{2}/2\Gamma}\ .
\end{equation}
As we will see, stochastic forcing also needs to be introduced to regulate strange attractors 
at small scales.   As the fractal structure of a strange attractor cannot be resolved on a lattice, it is necessary
to smooth the structure at the lattice length scale.  We use additive stochastic 
forcing for this purpose.  

Direct solution of the FPE is most commonly carried out for one dimensional systems.  
Extension to higher dimensions is conceptually straightforward, but numerically
challenging \cite{pichler2013numerical}.  The objective of the present paper is to apply the FPE to a three dimensional chaotic system. 
Numerical solutions have been developed
based on finite elements \cite{bergman1992robust,von2000calculation},
finite differences \cite{kumar2006}, and path integrals \cite{naess1994response}.  Here we depart from these traditional methods by instead directly solving for the zero or null mode of a finite-difference discretized FPE operator, {thus obviating the time-consuming and costly steps of computing the transient probability distributions. We illustrate this new method by applying it to the Lorenz-63 system \cite{lorenz1963deterministic} with additive stochastic forcing. Although a phenomenological FPE has been applied for a quantum system without the addition of stochastic forcing \cite{PhysRevLett.110.050401}, we follow previous work \cite{thuburn2005climate,gradivsek2000analysis,agarwal2016maximal,PhysRevE.92.062922,Kumar20142040} and add small additive white noise to wash out fractal
structure below the lattice scale.

Since numerical  solution of the FPE in larger numbers of dimensions is stymied by the ``curse of dimensionality'' it is important to develop
alternative forms of DSS.  Accordingly we also explore a second type of DSS,  an expansion in equal-time cumulants that can be applied to high-dimensional dynamical systems. A cumulant expansion was employed for the Orszag-McLaughlin attractor in Ref. \cite{ma2005exact}. For the Lorenz attractor
we show that low-order statistics are well reproduced at third order truncation.

The paper is organized as follows.  Section \ref{Lorenz} briefly describes the Lorenz-63 system with additive stochastic forcing, and its 
numerical integration.   Two different sets of parameters are considered, both of which yield chaotic behavior. 
Section \ref{FokkerPlanckEquation} describes the FPE and the numerical method that we use to find the invariant measure.  
Equal-time statistics so obtained are compared to those found through accumulation by DNS.  We study the scaling of the 
spectral gap of the linear FPE operator as the stochastic forcing is varied. 
Two extensions to the method are also considered: A self-adjoint construction of the linear operator, and the replacement of diffusion with hyperdiffusion.    
Section \ref{CE}  presents the cumulant expansion technique
and its evaluation by comparison to DNS.  Some conclusions are presented in Section \ref{Conclusion}.

\section{Lorenz-63 Attractor With Additive Stochastic Forcing}
\label{Lorenz}

The Lorenz-63 attractor is a three dimensional, chaotic system that was originally derived
by applying a severe Galerkin approximation to the equations of motion (EOMs) for Rayleigh-B\'{e}nard convection
with stress-free boundary conditions \cite{lorenz1963deterministic}.  We study an extension with additive
stochastic forcing  \cite{thuburn2005climate,gradivsek2000analysis,agarwal2016maximal,PhysRevE.92.062922,Kumar20142040}
that obeys Eq. \ref{stochastic}.  Such additive white noise can model fast or unresolved physical processes 
that are not explicitly described.
\begin{eqnarray}
\dot{x} &=& \sigma (y - x) + \eta_{1}(t)
\nonumber \\
\dot{y} &=& x (\rho-z) - y + \eta_{2}(t)
\nonumber \\
\dot{z} &=& x y - \beta z + \eta_{3}(t)
\label{EOM}
\end{eqnarray}
In this context $x$ is proportional to convective intensity, $y$ to the difference
in temperature between ascending and descending currents, and $z$
to the vertical temperature profile's deviation from linearity. 
Parameters $\beta$, $\rho$, and $\sigma$ are a geometric factor, 
the Rayleigh number, and the Prandtl number respectively.   We also choose the 
covariance of the Gaussian white noise to be diagonal and isotropic: $\Gamma_{ij} = \Gamma~ \delta_{ij}$.

We study the strange attractor at two different sets of parameters.
The conventional (classic) parameters are $\beta = 1.0$, $\rho = 26.5$, and $\sigma = 3.0$.
We also examine the case with the geometric factor changed to $\beta = 0.16$. 
This modification enhances layering in the PDF.  Since the geometric factor is related to the size of the attractor, a reduction in $\beta$ shrinks the attractor which in turn allows for a reduction in the stochastic forcing whilst keeping the structure sufficiently smooth at the lattice length scale.  Each attractor has a slow time scale
corresponding to transitions between the wings of the butterfly, and a fast time scale for
orbits about each wing.  The fast time-scale is $\tau \simeq 1$.

\subsection{Direct Numerical Simulation}
\label{DNS}

The Eq. \ref{EOM} EOMs are integrated forward in time with the fourth-order accurate Runge-Kutta 
algorithm with fixed time step $\delta t=0.01$.  Stochastic forcing, drawn from a normal distribution, is updated at intervals of $\Delta t = 0.1$ and interpolated at intermediate times $\delta t$.  Note that there is a good separation of time scales with $\tau \gg \Delta t \gg \delta t$ following the reasoning of Ref. \cite{lilly1969numerical}. 

Statistics are accumulated up to a final time $T = 2 \times10^{7}$.  The PDF is estimated from the histogram that results from binning the trajectory into cubic boxes.  The PDF is projected onto a plane by integrating over the direction perpendicular to the plane; for instance:
\begin{equation}
P(x, y) = \int_{-\infty}^\infty P(x, y, z)~ dz\ . 
\end{equation}
Figures \ref{ClassicParameters}, \ref{ModifiedParameters} and \ref{CrossSectionForcedUnforced}
show that even small stochastic forcing smoothes out the fine structure of the strange attractor \cite{Heninger2016}; 
in particular the ring-like steps in the PDF disappear.    

\begin{figure}[ht]
\includegraphics[width=\columnwidth]{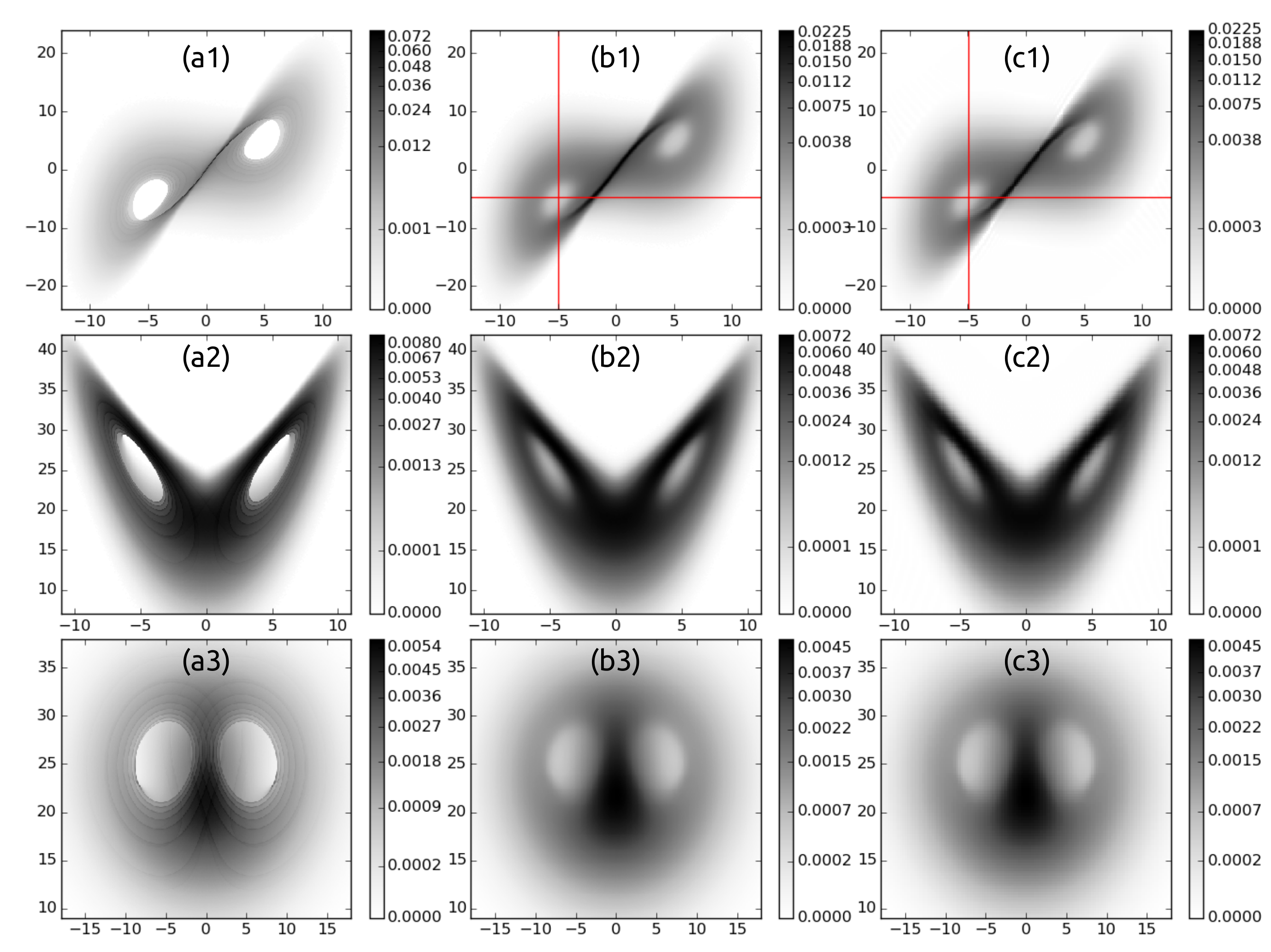}
\caption{(Color online) (a) PDF of the unforced Lorenz system (binned on a $637^{3}$ grid)
accumulated by DNS.  Classic Lorenz parameters are used (see Sec. \ref{Lorenz}).
(b) Same, but for the stochastically forced $(\Gamma=0.2)$ ($478^{3}$ grid).  
Note that the fine rings visible in the unforced system are 
washed out by the noise. 
(c) PDF of the stochastically forced attractor ($\Gamma=0.2$) as obtained from
the zero-mode of $\hat{L}_{FPE}$ on a $160^{3}$ grid and for $x \in [-12.5,~ 12.5]$,
$y \in [-24,~ 24]$ and $z \in [1,~ 45]$.  The JDQR algorithm is employed.  Rows (1), (2) and (3) correspond to the $x-y$,
$x-z$ and $y-z$ projections of the PDF respectively. Good agreement
between (b) DNS and (c) FPE is evident. Red lines correspond
to the cross-sections shown in Figure \ref{CrossSectionClassic}.}
\label{ClassicParameters}
\end{figure}

\section{Fokker-Planck Equation}
\label{FokkerPlanckEquation}

The FPE is an attractive alternative to the accumulation of statistics by DNS.  The linear FPE operator for the Lorenz
system,
\begin{equation}
\hat{L}_{FPE} P = \vec{\nabla} \cdot [(\sigma (y - x),~ x (\rho - z) -  y,~ x y - \beta z) P] - \Gamma \nabla^{2} P
\label{LorenzLFPE}
\end{equation}
is positive-definite in the sense that the real part of the eigenvalues are non-negative (if that were not the case, Eq. \ref{FPE} would diverge).   
The equal-time PDF of the FPE 
is the zero or null mode of $\hat{L}_{FPE}$.  By discretizing the operator on a lattice, the problem of finding the
zero mode is converted as a problem of sparse linear algebra.  

\subsection{Numerical Solution}
\label{NumericalSolution}

A standard center-difference scheme is used to discretize the derivatives that appear in Eq. \ref{LorenzLFPE}:   
\begin{eqnarray}
f_i^\prime &\approx& \frac{f_{i+1}-f_{i-1}}{2\Delta x}
\nonumber \\
f_i^{\prime \prime} &\approx& \frac{f_{i+1}-2f_{i}+f_{i-1}}{\Delta x^{2}}\ .
\label{FiniteDifference}
\end{eqnarray}
The PDF of the unforced Lorenz attractor has compact support but once stochastic forcing is included the PDF is non-zero but exponentially small
throughout phase space.  In the numerical calculation, the probability density $P$ is taken to vanish outside of the domain.
The zero-mode of $\hat{L}_{FPE}$ is found with the use of a preconditioned
Jacobi-Davidson QR (JDQR) algorithm \cite{sleijpen2000jacobi,fokkema1998jacobi}
that computes the partial Schur decomposition of a matrix with error tolerance set to $10^{-5}$.  We have checked that our results do 
not change appreciably for tighter tolerances.  Although the JDQR algorithm is well-suited 
for sparse matrices, requiring only the action of a matrix multiplying a vector, we were unable
to find a suitable sparse preconditioner.  A sparse preconditioner
would enable a substantial increase in resolution. The Jacobi correction equation is preconditioned following section 3.2 of Ref. \cite{fokkema1998jacobi} and solved using the generalized minimal residual (GMRES) method using Matlab.  The highest resolution we have been
able to reach is $160^{3}$ with 500 GB of memory.  To the best of our knowledge this is the largest number of grid points for which the Fokker-Planck equation has been simulated (for comparison, more than double the number reached by Kumar and Narayanan (2016) \cite{kumar2006}). This is made possible by re-expressing the time-dependant Fokker-Planck problem to a sparse linear algebra problem.  On the $160^{3}$ grid the $\hat{L}_{FPE}$ matrix has $28,518,400$ non-zero elements and a sparsity of $1.70 \times 10^{-6}$. 

Modes at the maximal wavenumber decay diffusively due to the stochastic forcing 
with a decay timescale $\tau_d$ given by
\begin{equation}
\tau_d^{-1} =  \frac{\pi^2 \Gamma}{4} [ (\Delta x)^{-2} + (\Delta y)^{-2} + (\Delta z)^{-2} ]
\label{decayRate}
\end{equation}
Equating this time scale to the timescale of fast dynamics $\tau$ provides an estimate of the 
effective resolution of the discretized FPE.

\begin{figure}[ht]
\includegraphics[width=\columnwidth]{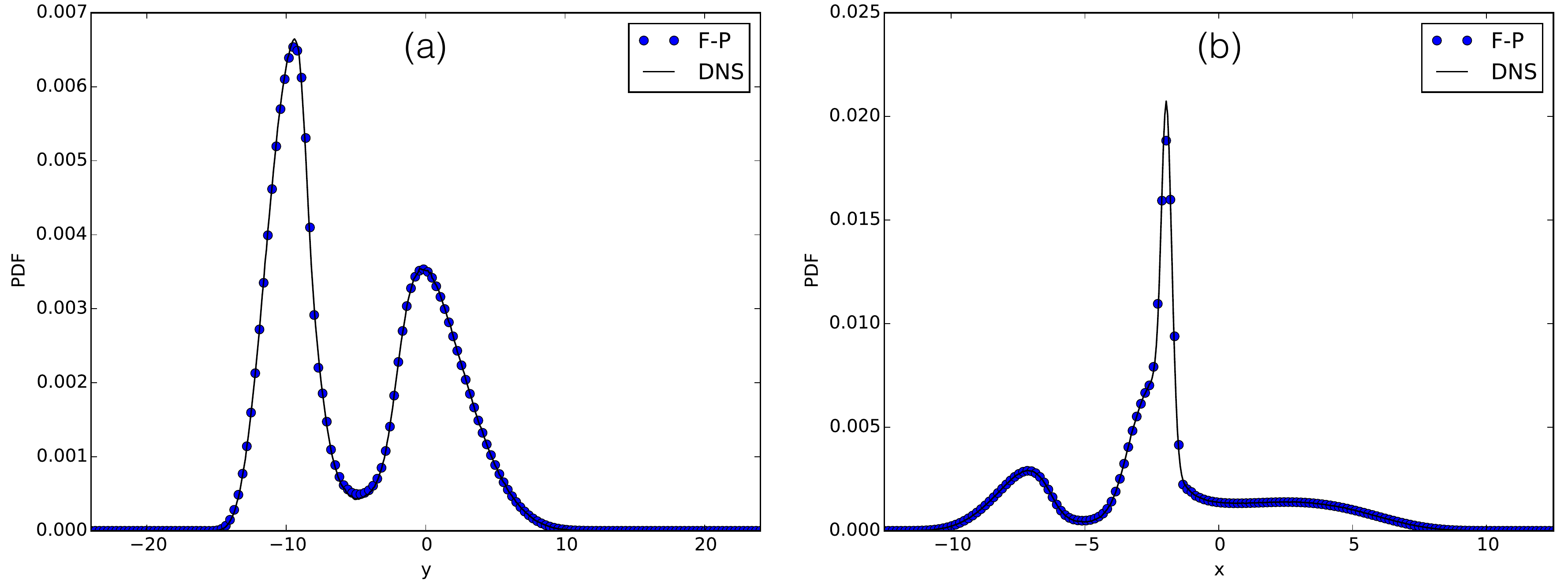}
\caption{(a) Vertical and (b) horizontal slices along the red lines of the
$x-y$ planar projection shown in Fig. \ref{ClassicParameters}.  
The DNS and FPE methods agree quantitatively.}
\label{CrossSectionClassic}
\end{figure}

\begin{figure}[ht]
\includegraphics[width=\columnwidth]{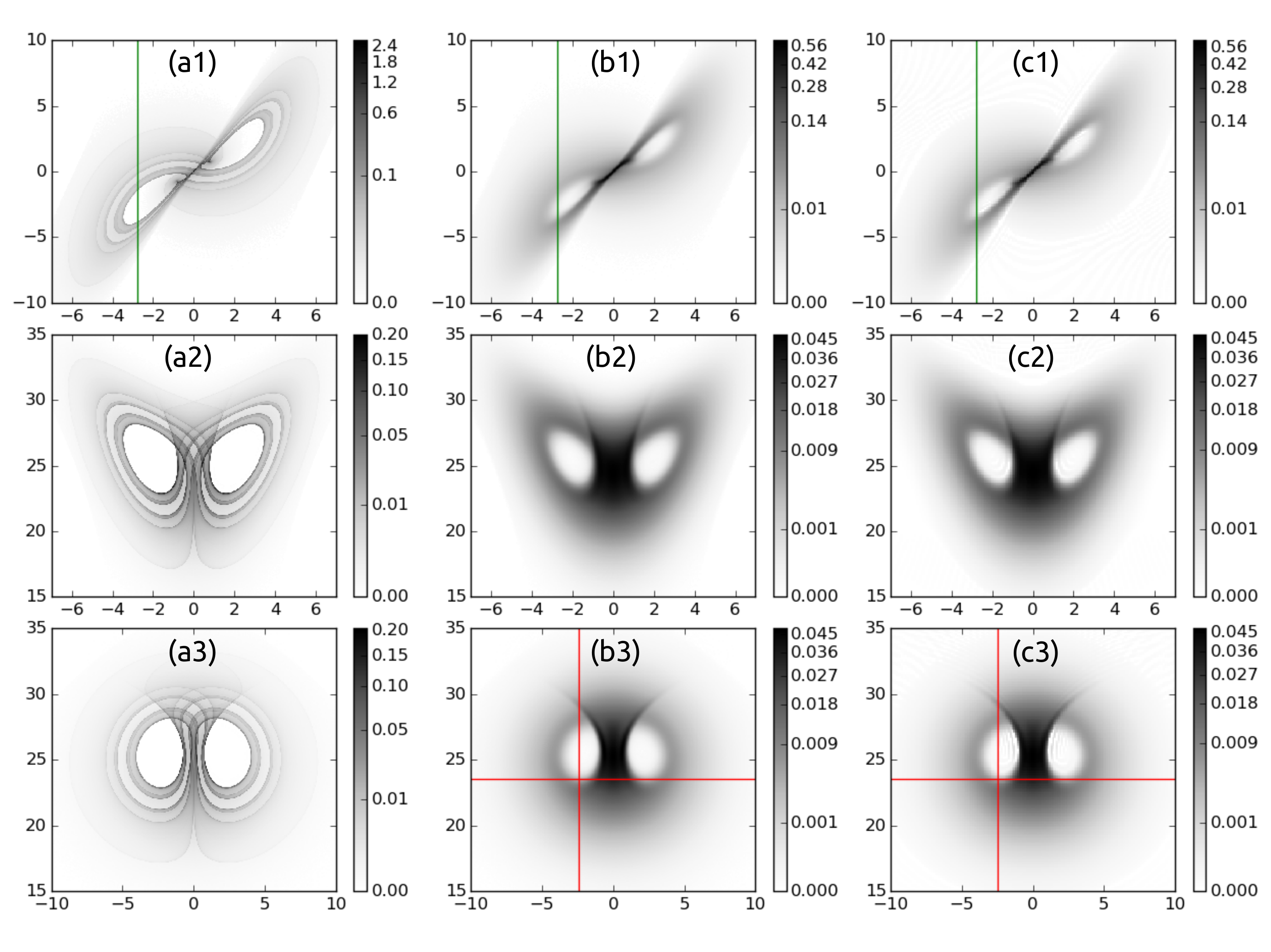}
\caption{(Color online) (a) PDF of the unforced Lorenz system (binned on a $637^{3}$ grid)
accumulated by DNS.  Modified parameters are used (see Sec. \ref{Lorenz}).
(b) Same, but for the stochastically forced $(\Gamma=0.02)$ ($478^{3}$ grid).  
Note that the fine rings visible in the unforced system are 
washed out by the noise. 
(c) PDF of the stochastically forced attractor ($\Gamma=0.02$) as obtained from
the zero-mode of $\hat{L}_{FPE}$ on a $160^{3}$ grid and for $x \in [-7,~ 7]$,
$y \in [-10,~ 10]$ and $z \in [15,~ 35]$.   Rows (1), (2) and (3) correspond to the $x-y$,
$x-z$ and $y-z$ projections of the PDF respectively. Good agreement
between (b) DNS and (c) FPE is evident. Red lines correspond
to the cross-sections shown in Figure \ref{CrossSectionModified} and green lines correspond
to the cross-section shown in Figure \ref{CrossSectionForcedUnforced}}.
\label{ModifiedParameters}
\end{figure}

\begin{figure}[ht]
\includegraphics[clip,width=\columnwidth]{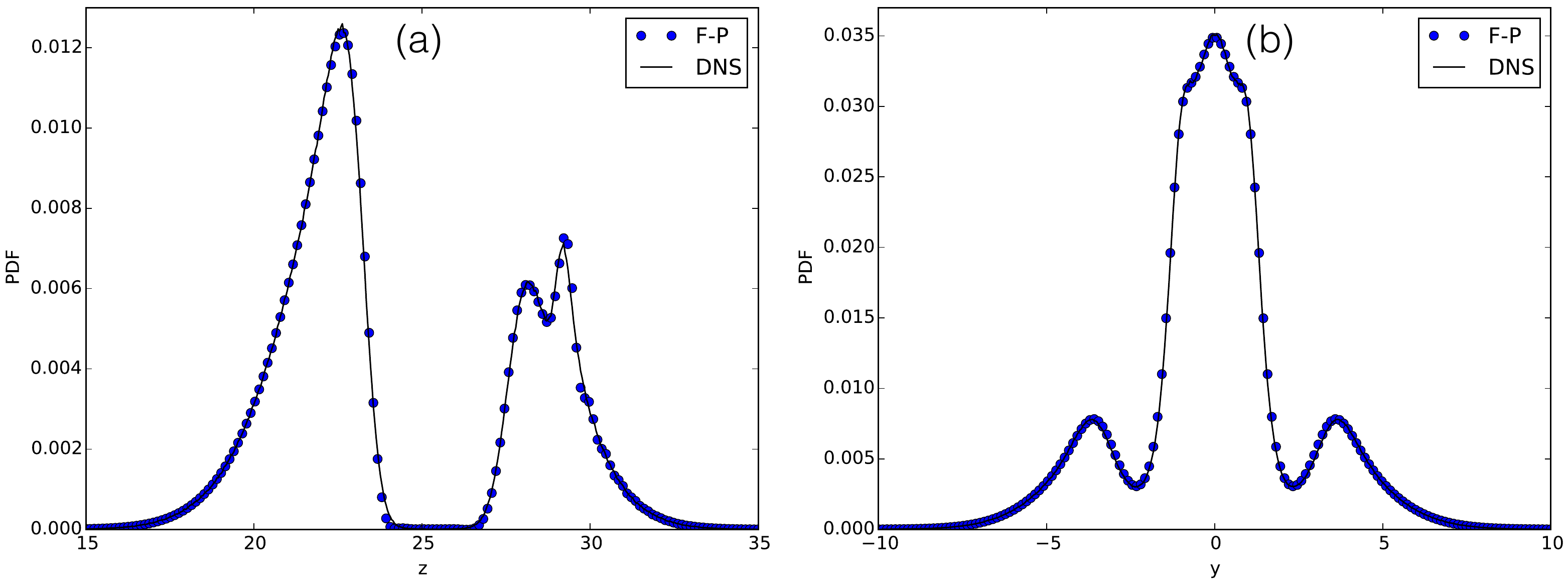}
\caption{(a) Vertical and (b) horizontal slices along the red lines of the
$y-z$ planar projection shown in Fig. \ref{ModifiedParameters}.  
The DNS and FPE methods agree quantitatively.}
\label{CrossSectionModified}
\end{figure}

\begin{figure}[ht]
\includegraphics[width=0.5\columnwidth]{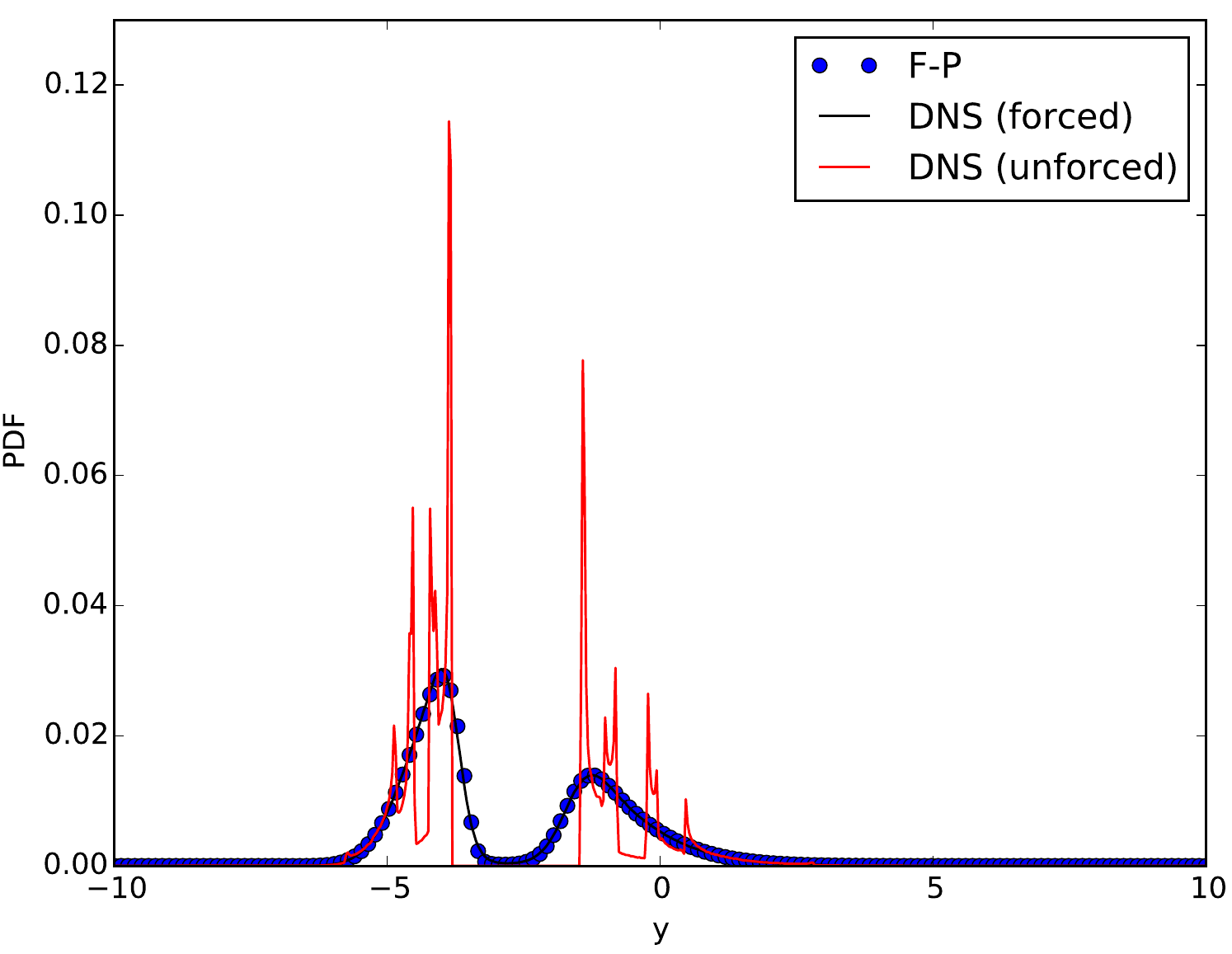}
\caption{(Color online) Cross-section through the $x-y$ planar projection of the PDF where
$x=-2.77$ for unforced DNS, stochastically forced DNS, and stochastically forced
FPE (with $\Gamma=0.02$). Modified parameters are used. Sharp peaks in probability
seen in the unforced Lorenz system are eliminated by the stochastic forcing.}
\label{CrossSectionForcedUnforced}
\end{figure}

Figures \ref{ClassicParameters}, \ref{CrossSectionClassic}, \ref{ModifiedParameters}, \ref{CrossSectionModified} and \ref{CrossSectionForcedUnforced}
show good quantitative agreement between PDFs accumulated by DNS and those obtained from the 
zero mode of the FPE operator.  Fine structure evident in the $x-z$ and $y-z$ projections of the PDF for the Lorenz
attractor with modified parameters, Figure \ref{ModifiedParameters}, is reproduced by the FPE. 
The modified parameters also demonstrate the sensitivity of the fractal
structure to stochastic forcing.  Even forcing as small as $\Gamma=0.02$ washes out the 
ring-like steps in the PDF in Figure \ref{ModifiedParameters}, also shown by the selected horizonal slice in Figure \ref{CrossSectionForcedUnforced}.  

\subsection{Eigenvalue Spectra of the Fokker-Planck Operator}
\label{Gap}

The gap to the eigenmode with the smallest non-zero real part of its eigenvalue is of interest as it corresponds to the slowest relaxation rate in the statistics towards the statistical steady state.   Likewise the imaginary components of the eigenvalues of the linear FPE operator set the quasi-periodic frequency of the dynamical system (see Section \ref{Lorenz} for a discussion of the slow and fast time scales).  This interpretation of the real and imaginary components of the eigenvalues can be illustrated analytically with a two-dimensional linear Ornstein-Uhlenbeck system with circular orbits of period $2 \pi$ that decay at rate $\alpha$:
\begin{eqnarray}
\dot{x} & = & y - \alpha x + \eta_1(t)
\nonumber \\
\dot{y} & = & -x - \alpha y + \eta_2(t).
\label{2DCircular}
\end{eqnarray}
The eigenvalues of the FPE operator of this system are $\lambda_{n,m} = \alpha n \pm im$
where $m,~ n$ are non-negative integers that are either both odd, or both even, and $m \leq n$.  (The eigenvalues can be found by transforming to polar coordinates and expressing the PDF as a separable solution $P(r,~ \theta)=R(r)~ \Theta(\theta)$ with the radial equation solved using generalized Laguerre polynomials.)  Because the system of Eq. \ref{2DCircular} is linear, the stochastic forcing does not affect the spectra.  The real parts of the eigenvalues are determined by the decay rate $\alpha$ whereas the imaginary part is set by the orbital period. 

The spectrum of low-lying eigenvalues is shown in Figure \ref{EnergyVersusGamma} with the complex eigenvalues appearing in conjugate pairs because $\hat{L}_{FPE}$ is purely real-valued. Increased stochastic forcing increases the values of the real components of the eigenvalues of the FPE operator, in turn shortening the relaxation timescale.  Linear extrapolation to the unforced 
$(\Gamma=0)$ limit yields a real eigenvalue of $0.0802$ corresponding to a relaxation timescale of about $12$ in accord with expectations.
\begin{figure}[ht]
\includegraphics[width=1.0\columnwidth]{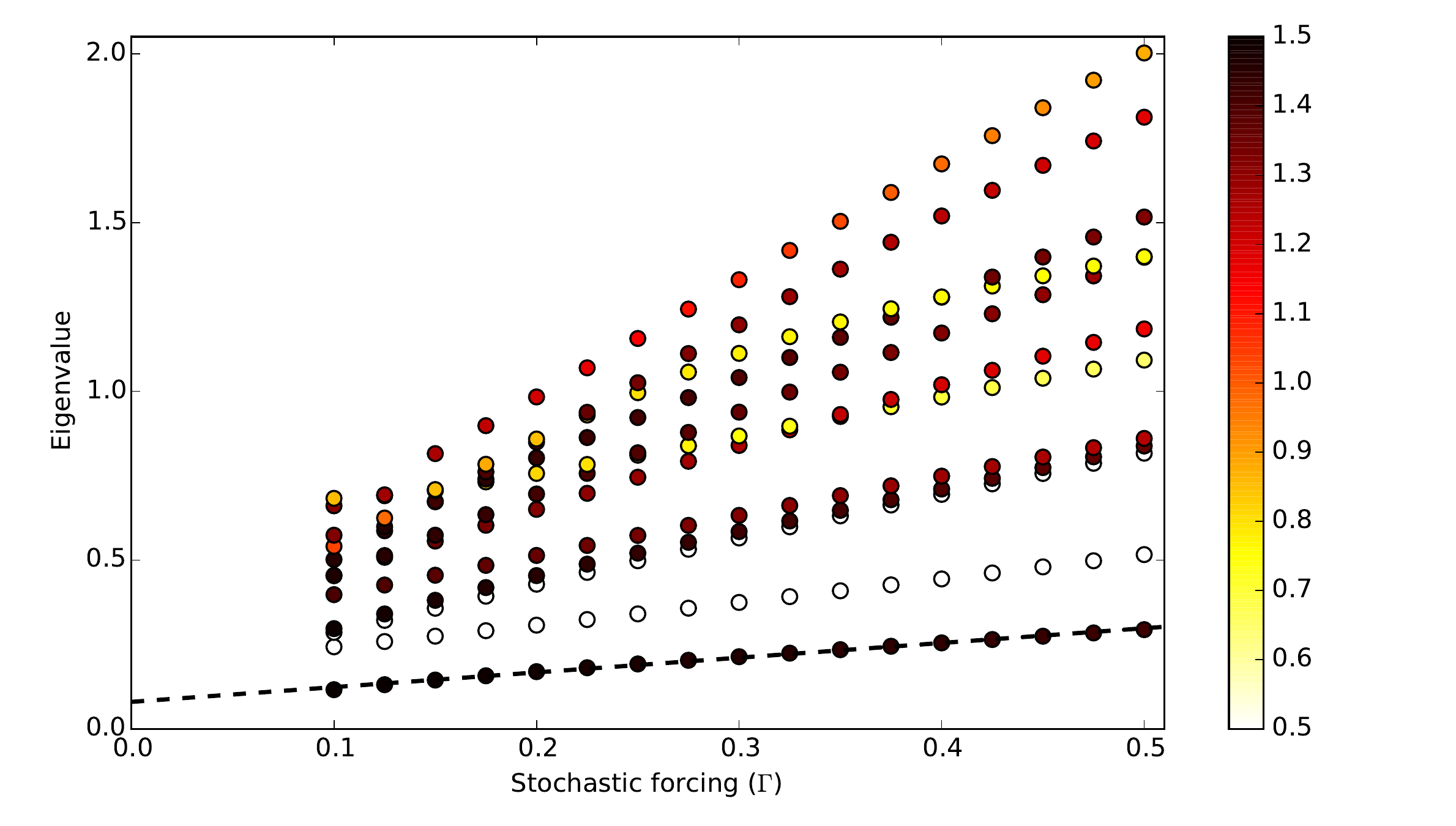}
\caption{(Color online) Real component of the first 12 non-zero eigenvalues of the
linear FPE operator of the Lorenz system (on an $80^3$ lattice
grid) as a function of stochastic forcing $(\Gamma)$. Modified parameters
are used (see Sec. \ref{Lorenz}) with $x \in [-12.5,~12.5]$, $y \in [-17.5,~17.5]$
and $z \in [7.5,~ 42.5]$. The phase of each eigenvalue is represented
by the color of the points with white correspond to a phase of zero.  
A linear extrapolation of the first non-zero eigenvalue to $\Gamma = 0$ is shown.  The second and third non-zero eigenvalues are purely real.}
\label{EnergyVersusGamma}
\end{figure}
Likewise the non-zero imaginary parts of the eigenvalues in Fig. \ref{EnergyVersusGamma} correspond to the oscillation frequency around each of the two wings of the Lorenz attractor as well as the slower rate of transitions between the two wings.

\subsection{Extensions}
\label{Extensions}

We consider two modifications to the FPE operator.   Self-adjoint constructions of the linear operator are considered first, and then the replacement of diffusion with hyperdiffusion. 

\subsubsection{Self-Adjoint Linear FPE Operators:}
\label{Hermitian}

We investigate whether or not self-adjoint generalizations of the linear FPE operator have any advantages.   One way to construct such an operator
is to double the size of the linear space by introducing the operator 
\begin{equation}
\hat{H}_2 =
\left( \begin{array}{cc}
0 & \hat{L}_{FPE} \\
\hat{L}_{FPE}^\dagger & 0
\end{array} \right)\
\label{H2}
\end{equation}
that by construction obeys $\hat{H}_2^\dagger = \hat{H}_2$.  $\hat{H}_2$ is no longer positive-definite; 
instead its eigenvalue spectrum is symmetric about 0.  Finding the zero mode therefore 
requires a numerical algorithm that can target the middle of the spectrum.  
Alternatively the operator
\begin{equation} 
\hat{H}_1 \equiv \hat{L}_{FPE}^\dagger~ \hat{L}_{FPE}
\label{H1}
\end{equation}
is self-adjoint, positive-definite, and has the same ground state as $\hat{L}_{FPE}$.  
To see this define $| V \rangle \equiv \hat{L}_{FPE} | \Psi_{0} \rangle$ where $| \Psi_{0}\rangle$
is the zero mode of $\hat{H}_1$. Then $\langle \Psi_{0} | \hat{L}_{FPE}^{\dagger} \hat{L}_{FPE} | \Psi_{0} \rangle = \langle V | V \rangle = 0$
that implies that $|V \rangle$ must have zero norm, $| V \rangle = | 0 \rangle$, and $| \Psi_{0} \rangle$ is the zero mode of $\hat{L}_{FPE}$.   The ground
state of a quantum system has (under rather general conditions) no nodes so $\Psi_0(\vec{x})$ automatically produces a non-negative PDF -- an added
virtue of viewing steady-state solutions of the FPE as a problem of linear algebra.

Because the ground state of $\hat{H}_1$ might possibly be found using simulated annealing or quantum annealing / adiabatic quantum computation  \cite{finnila1994quantum,santoro2006optimization}, it 
is of interest to investigate its numerical solution.  A drawback of $\hat{H}_1$ is that the eigenvalues are clustered more tightly around $0$ than 
those of $\hat{L}$, slowing convergence.  
We find that the ground state, computed using the Davidson algorithm \cite{davidson1975iterative,davidson1993monster}, matches that obtained from $ \hat{L}_{FPE}$ alone.

\subsubsection{Hyperdiffusion:}
\label{Hyperdiffusion}

Another interesting variant to investigate is to replace diffusion in 
Eq. \ref{LorenzLFPE} with biharmonic hyperdiffusion:
\begin{equation}
-\Gamma \nabla^{2} \rightarrow  \Gamma_2 \nabla^4
\label{hyperdiffusion}
\end{equation}
As hyperdiffusion is more scale selective than ordinary diffusion, it acts to smooth
out structure at the lattice scale while leaving larger structures intact.  
Holding the dissipation rate at the lattice scale fixed, $\Gamma_2$ scales inversely with the 4th power of the spacing (contrast with Eq. \ref{decayRate}).  Unlike 
ordinary diffusion that has a physical origin in stochastic forcing, hyperdiffusion violates 
realizability leading to negative probability densities (see for example section 4.3 of Ref. \cite{risken1996fokker}).   Also hyperdiffusion reduces 
the sparseness of the discretized linear FPE operator, slowing computation.  On an $80^3$ grid the PDF shows less fine structure as expected.  

\section{Cumulant Expansion}
\label{CE}

Numerical solution of the FPE for dynamical systems of dimension greater than 3 is increasingly difficult.  Therefore it is of 
interest to explore other approaches to DSS.   Here we use an expansion in low-order equal-time cumulants to find statistics of the Lorenz attractor.
In addition to the Orszag-McLaughlin attractor \cite{ma2005exact}, it has also been applied to a number of 
high-dimensional problems in fluids using spatial averaging \cite{marstonetal2008,tobiasdagonetal2011,tobiasmarston2013,parkerkrommes2013a,parkerkrommes2014,squirebhattacharjee2014,marston2014direct,aitchaalalmarstonetal2016} although sometimes ensemble averages are employed \cite{Bakas:2013bo,Bakas:2015iy}.

The EOMs for equal-time cumulants can be elegantly derived using the Hopf functional formalism \cite{frisch1995,tobiasdagonetal2011}.  However
it is also possible to derive them directly from the EOMs of the dynamical system.  Consider a general system with quadratic nonlinearities: 
\begin{equation}
\dot{q}_i = F_i + L_{ij}~ q_{j} + Q_{ijk}~ q_{j} q_{k} + \eta_i.
\label{QuadraticEOM}
\end{equation}
A Reynolds decomposition, 
\begin{equation}
q_{i} = \overline{q_{i}} + q_{i}^\prime\ ,
\end{equation}
of the dynamical variables $q_i$ into a mean and a fluctuation is the starting point.  Here (and as discussed below) the average is chosen to be a mean over initial conditions drawn from a Gaussian distribution. The Reynolds decomposition obeys the rules:
\begin{eqnarray}
\overline{\overline{q}_i} &=& \overline{q}_i
\nonumber \\
\overline{q^\prime_i} &=& 0
\nonumber \\
\overline{\overline{q}_i~ q_j} &=& \overline{q}_i~ \overline{q}_j.
\end{eqnarray}
The first cumulant, $c_{i}$, is the mean of $q_{i}$ and the second and third cumulants 
are given by centered moments of the fluctuations.  By contrast the fourth cumulant is {\it not} a centered moment:  
\begin{eqnarray}
c_{i} &\equiv& \overline{q_{i}}
\nonumber \\
c_{ij} &\equiv& \overline{q_{i}^\prime q_{j}^\prime}
\nonumber \\
c_{ijk} &\equiv& \overline{q_{i}^\prime q_{j}^\prime q_{k}^\prime}
\nonumber \\
c_{ijk\ell} &\equiv& \overline{q_{i}^\prime q_{j}^\prime q_{k}^\prime q_{\ell}^\prime}  - c_{ij}~ c_{k\ell} - c_{ik}~ c_{j\ell} - c_{i\ell}~ c_{j k}\ .
\label{cumulants}
\end{eqnarray}
These definitions ensure that the third and higher cumulants of a Gaussian distribution vanish.

The EOMs for the cumulants may be found directly.  The equation of motion for the
first cumulant is obtained by ensemble averaging Eq. \ref{QuadraticEOM} and using the definitions of
the first and second cumulants in Eq. \ref{cumulants},
\begin{eqnarray}
\frac{dc_{i}}{dt} & = & \overline{\frac{dq_{i}}{dt}}\nonumber \\
& = & F_i + L_{ij}~ \overline{q_{j}} + Q_{ijk}~ \overline{q_{j} q_{k}}\nonumber \\
& = & F_i + L_{ij}~ c_{j} + Q_{ijk}~ (c_{j} c_{k}+c_{jk}).
\label{CE1}
\end{eqnarray}
The tendency of the first cumulant involves the second cumulant.  Truncating at first order by discarding the second 
cumulant, the CE1 approximation, shows that the first cumulant obeys the same EOMs as Eq. \ref{QuadraticEOM}; 
hence there is no non-trivial fixed point solution.  In general for quadratically nonlinear systems the tendency of
the $(n)$-th cumulant requires knowledge of the $(n+1)$-th cumulant, the well-known closure problem.  
Thus the EOM for the second cumulants requires the first, second and third cumulants:
\begin{eqnarray}
\frac{dc_{ij}}{dt} & = & 2 \left\{ \overline{\frac{dq_{i}^\prime}{dt}~ q_{j}^\prime} \right\} \nonumber \\
& = & 2 \left\{ \overline{\left(\frac{dq_{i}}{dt} - \frac{d\overline{q_{i}}}{dt}\right) q_{j}^\prime} \right\} \nonumber \\
& = & 2 \left\{ \overline{\frac{dq_{i}}{dt} q_{j}^\prime}\right\} \nonumber \\
& = & 2 \left\{ \overline{L_{ik}~ q_{k} q_{j}^\prime + Q_{ik\ell}~ q_{k}q_{\ell}q_{j}^\prime} \right\}  + 2 \Gamma_{ij} \nonumber \\
& = & \left\{ 2L_{ik}~ c_{kj} + Q_{ik\ell}~ (4 c_{k} c_{\ell j} + 2c_{k \ell j})\right\} + 2 \Gamma_{ij}\ .
\label{CE2}
\end{eqnarray}
The covariance matrix $\Gamma_{ij}$ of the stochastic forcing appears in the last two lines as (implicitly) a short-time averaging
has been carried out in addition to ensemble averaging.  The symmetrization operation $\left\{ \ldots \right\}$ over all permutations of the free indices has been 
introduced for conciseness. In the case of a $2$-index variable such as the second cumulant it is
defined as $\left\{ c_{ij}\right\} =\frac{1}{2}(c_{ij}+c_{ji})$ and
similarly for higher cumulants.

Because each higher cumulant carries an additional dimension with it, closure
should be performed as soon as possible. Closing the EOMs at second order by 
discarding the contribution of the third cumulant $c_{ijk}$ (CE2) is sometimes possible \cite{marstonetal2008}
and yields a realizable approximation (because the PDF is Gaussian that is non-negative everywhere).  The CE2 EOMs 
can then be integrated forward in time by specifying as an initial condition non-zero first and second cumulants, corresponding to a 
normally distributed initial ensemble.  
However the Lorenz attractor is so nonlinear that the CE2 EOMs, integrated forward in time,
do not reach a fixed point that would characterize a statistical steady state.  Therefore we proceed to next order, CE3,
by setting the fourth cumulant to zero, $c_{ijkl}=0$. The fourth centered moment may then be 
expressed in terms of the second and third cumulants: 
\begin{eqnarray}
\overline{q_{m}^\prime q_{n}^\prime q_{j}^\prime q_{k}^\prime} &=& c_{mn} c_{jk} + c_{mj} c_{nk} + c_{mk} c_{jn} + c_{mnjk} \nonumber \\
&\simeq& c_{mn}c_{jk} + c_{mj} c_{nk} + c_{mk} c_{jn}\ \ \ {\rm (CE3)}
\end{eqnarray}
and the EOM for the third cumulant now closes:
\begin{eqnarray}
\frac{dc_{ijk}}{dt} & = & 3\left\{ \overline{\frac{dq_{i}^\prime}{dt} q_{j}^\prime q_{k}^\prime} \right\} 
\nonumber \\
&=& 3 \left\{ \overline{\left(\frac{dq_{i}}{dt} - \frac{d\overline{q_{i}}}{dt}\right) q_{j}^\prime q_{k}^\prime} \right\} 
\nonumber \\
&=& 3 \left\{ \overline{\frac{dq_{i}}{dt} q_{j}^\prime q_{k}^\prime} - \frac{dc_{i}}{dt} \overline{q_{j}^\prime q_{k}^\prime} \right\} 
\nonumber \\
&=& 3 \left\{ L_{im}~ c_{mjk} + Q_{imn}~ \left( 2 c_{m} c_{njk} - c_{mn} c_{jk} + \overline{q_{m}^\prime q_{n}^\prime q_{j}^\prime q_{k}^\prime} \right) \right\}
\nonumber \\
&=& \left\{ 3L_{im}~ c_{mjk} + 6Q_{imn}~ \left( c_{m} c_{njk} + c_{mj} c_{nk}\right) \right\} - \frac{c_{ijk}}{\tau_{d}} 
\label{CE3}.
\end{eqnarray}
In the final line of Eq. \ref{CE3} a phenomenological eddy damping time scale, $\tau_{d}$, has been introduced to model
the neglect of the fourth cumulant \cite{orszag1974lectures,marston2014direct}.  CE2 is recovered in the limit $\tau_d \rightarrow 0$ as the 
third cumulant is suppressed in this limit. As $\tau_d$ is increased, contributions from the interactions of two fluctuations that produce another fluctuation begin to be felt. These nonlinear {\it fluctuation + fluctuation $\rightarrow$ fluctuation} interactions are dropped at order CE2, which retains only the {\it fluctuation + fluctuation $\rightarrow$ mean}  and {\it fluctuation + mean $\rightarrow$ fluctuation} interactions \cite{marston2014direct}.  This can be seen be examining Eqs. \ref{CE1} and \ref{CE2} and observing that the second cumulant only interacts through $Q_{ijk}$ with the first cumulant, and not with itself.  
Upon increasing $\tau_d$ further eventually realizability is lost as the second cumulant develops 
negative eigenvalues that are unphysical 
\cite{kraichnan1980realizability,hanggi1980remark,marston2014direct}.
Furthermore the appearance of a negative eigenvalue triggers an instability and the time-integrated EOMs diverge.  We note that 
non-zero third cumulants are a mark of a non-Gaussian PDF for which all higher cumulants would generically be non-zero as well.  
Therefore it is inconsistent to discard the fourth and higher cumulants, and that inconsistency makes itself felt in non-realizability.  

An appealing feature of cumulant expansions is that they respect the symmetries of the dynamical
system.  For instance, the CE3 fixed point exactly respects $\overline{x} = \overline{y} = 0$ due to the invariance
of the Lorenz system under the reflection $\{x,y,z\}\rightarrow\{-x,-y,z\}$.  This statistical
symmetry is only approximately obeyed by DNS statistics accumulated over a finite time.
Table \ref{UnforcedCumulants} compares the first and second cumulants of the unforced Lorenz attractor
as accumulated by DNS to those obtained from CE3 for two different choices of the time scale $\tau_d$.  
Good qualitative agreement is found for both $\tau_d = 0.1$ and $\tau_d = 0.5$ demonstrating an
insensitivity to the precise choice of the time scale.  Table \ref{ForcedCumulants} displays the same cumulants, 
but now for the stochastically forced attractor.  In addition statistics obtained from the FPE zero mode are shown,   
calculated as the PDF-weighted sums of the desired variable over the domain of the lattice.  Though the small stochastic
forcing has a large effect on the fine structure of the PDF, it only changes the covariances slightly.

\begin{longtable}{|c|c|c|c|}
\caption{Comparison of the low-order statistically steady state cumulants of the unforced Lorenz attractor ($\Gamma = 0$) 
as accumulated by DNS and as calculated with CE3.  Modified Lorenz parameters are used (see Sec. \ref{Lorenz}).  Entries such as ${\cal O}(10^{-4})$
indicate that these mean values are tending to zero as time-averaging of DNS extends over increasing intervals of time.}
\endfirsthead
\hline
\label{UnforcedCumulants}
Cumulant & DNS & CE3 $(\tau_d=0.1)$ & CE3 $(\tau_{d}=0.5)$\tabularnewline
\hline
\hline
$\overline{x}$ & ${\cal O}(10^{-4})$ & 0 & 0\tabularnewline
\hline
$\overline{y}$ & ${\cal O}(10^{-4})$ & 0 & 0\tabularnewline
\hline
$\overline{z}$ & 24.796 & 25.188 & 25.000\tabularnewline
\hline
$\overline{x^\prime x^\prime}$ & 3.966 & 4.030 & 4.000\tabularnewline
\hline
$\overline{x^\prime y^\prime}$ & 3.966 & 4.030 & 4.000\tabularnewline
\hline
$\overline{x^\prime z^\prime}$ & ${\cal O}(10^{-5})$ & 0 & 0\tabularnewline
\hline
$\overline{y^\prime y^\prime}$ & 5.395 & 4.392 & 4.908\tabularnewline
\hline
$\overline{y^\prime z^\prime}$ & ${\cal O}(10^{-5})$ & 0 & 0\tabularnewline
\hline
$\overline{z^\prime z^\prime}$ & 8.513 & 5.592 & 6.825\tabularnewline
\hline
\end{longtable}

\begin{longtable}{|c|c|c|c|c|}
\caption{As in Table \ref{UnforcedCumulants} but with added stochastic forcing ($\Gamma=0.02$).}
\endfirsthead
\hline
\label{ForcedCumulants}
Cumulant & DNS & FPE & CE3 $(\tau_d=0.1)$ & CE3 $(\tau_{d}=0.5)$\tabularnewline
\hline
\hline
$\overline{x}$ & ${\cal O}(10^{-5})$ & ${\cal O}(10^{-9})$ & 0 & 0\tabularnewline
\hline
$\overline{y}$ & ${\cal O}(10^{-5})$ & ${\cal O}(10^{-9})$ & 0 & 0\tabularnewline
\hline
$\overline{z}$ & 24.834 & 24.834 & 25.192 & 25.004\tabularnewline
\hline
$\overline{x^\prime x^\prime}$ & 3.977 & 3.978 & 4.034 & 4.004\tabularnewline
\hline
$\overline{x^\prime y^\prime}$ & 3.971 & 3.972 & 4.031 & 4.001\tabularnewline
\hline
$\overline{x^\prime z^\prime}$ & ${\cal O}(10^{-5})$ & ${\cal O}(10^{-8})$ & 0 & 0\tabularnewline
\hline
$\overline{y^\prime y^\prime}$ & 5.350 & 5.349 & 4.396 & 4.912\tabularnewline
\hline
$\overline{y^\prime z^\prime}$ & ${\cal O}(10^{-4})$ & ${\cal O}(10^{-8})$ & 0 & 0\tabularnewline
\hline
$\overline{z^\prime z^\prime}$ & 8.150 & 8.135 & 5.610 & 6.829\tabularnewline
\hline
\end{longtable}

\section{Conclusion}
\label{Conclusion}

Direct Statistical Simulation (DSS) is an attractive alternative to the accumulation of statistics by Direct Numerical Simulation (DNS). 
Transforming the problem of finding the equal-time statistics of dynamical systems into a problem of sparse linear algebra offers an accurate and elegant alternative to traditional approaches.  It would be interesting to employ a sparse preconditioner for the large non-symmetric FPE operator studied here as this should permit even
higher resolutions to be reached, possibly revealing the fine ring-like steps 
in the Lorenz attractor PDF.  Galerkin discretizations of the linear operator may also be interesting to explore.  

We also show that an expansion in equal-time cumulants through third order (CE3) is able to reproduce the low-order statistics of the
attractor, despite its highly nonlinear nature.  The cumulant expansion technique is especially good for higher dimensional systems due to its speed. Deterministic chaos and stochastic noise are seen to have similar effects on the low-order statistics \cite{knobloch1979statistical,agarwal2016maximal} with both contributing to the variance.  By contrast Figure \ref{CrossSectionForcedUnforced} shows that the deterministic dynamics of the strange attractor produces high-order statistics that stochastic forcing erases.

\section{Acknowledgments}
We are grateful to P. Zucker and D. Venturi for useful discussions.  This research was supported in
part by NSF DMR-1306806 and NSF CCF-1048701.

\bibliographystyle{unsrt}

\end{document}